\documentclass[10pt, conference, letterpaper]{IEEEtran}

\usepackage{amsmath, amssymb}

\ifCLASSINFOpdf
  \usepackage[pdftex]{graphicx}
  \usepackage{subfigure}
   \usepackage{tabularx}
  \graphicspath{{./figure/}}
  \usepackage{multirow}
  \usepackage{color}
  \usepackage{cite}
  \usepackage{algorithm}
  \usepackage{algpseudocode}
  \usepackage{threeparttable}
\else
\fi

\begin{document}
\title{Towards Cognitive Routing based on Deep Reinforcement Learning}

\author{\IEEEauthorblockN{Jiawei Wu, Jianxue Li, Yang Xiao, Jun Liu}
\IEEEauthorblockA{Center for Data Science\\
Beijing University of Posts and Telecommunications, Beijing, China, 100876.\\
Email: cloudsae, lijianxue, zackxy, liujun@bupt.edu.cn}
}

\maketitle

\begin{abstract}

Routing is one of the key functions for stable operation of network infrastructure. Nowadays, the rapid growth of network traffic volume and changing of service requirements call for more intelligent routing methods than before. Towards this end, we propose a definition of cognitive routing and an implementation approach based on Deep Reinforcement Learning (DRL). To facilitate the research of DRL-based cognitive routing, we introduce a simulator named RL4Net for DRL-based routing algorithm development and simulation. Then, we design and implement a DDPG-based routing algorithm. The simulation results on an example network topology show that the DDPG-based routing algorithm achieves better performance than OSPF and random weight algorithms. It demonstrate the preliminary feasibility and potential advantage of cognitive routing for future network.

\end{abstract}


\IEEEpeerreviewmaketitle

\section{Introduction}

Routing, the process of selecting a path for packet transmission in networks, is the key function for stable operation of network infrastructure. Basically, we can classify routing technologies into two categories, non-quality-aware and quality-aware. Most widely used routing protocols and algorithms, such as RIP\cite{RIP}, IGRP\cite{IGRP} and OSPF\cite{OSPF}, are non-quality-ware because they cannot make routing decision using network and service quality information. Although non-quality-aware routing protocols and algorithms are simple to be implemented on routers and have been worked well for many years, they are challenged by rapid growth of network traffic volume and changing of service requirements. Therefore, a number of quality-aware routing protocols and algorithms are proposed in recent years, which aim to choose paths with better performance by leveraging network quality metrics like delay, jitter and loss \cite{Pragyansmita2002, Hanzo2007, Chen2007}. However, they are not widely used because of higher requirement of computation capabity on routers and expensive upgrade cost.

In recent years, with the rapid progress of new technologies like SDN and NFV, a number of research works introduced that a good opportunity has been raised to implement more complex routing decision on powerful hardwares \cite{Xie2018}. For example, Google has proved the separation of routing control and operation approach is feasible for achieving better quality assurance on software defined networks \cite{Google}. Inspired by these works, we propose the concept of \textbf{cognitive routing} that extends the concept of quality-aware routing by introduce three key capabilities into the routing decision component, \textbf{inference}, \textbf{decision} and \textbf{learning}. Not just introduce the concept, we propose an implementation approach based on Deep Reinforcement Learning (DRL). To facilitate the research of DRL-based cognitive routing, we develop a simulator named \textbf{RL4Net} for DRL-based routing algorithm development and simulation. In addition, we design and implement a Deep Deterministic Policy Gradient (DDPG) based routing algorithm. To demonstrate the preliminary feasibility and potential advantage of cognitive routing, we compare the DDPG-based routing algorithm with OSPF and random weight algorithms. The simulation results on an example network topology show that the DDPG-based routing algorithm achieves better performance.

In summary, the main contributions of our paper are as follows:

\begin{itemize}
\item We introduce the concept of cognitive routing with an implementation approach based on deep reinforcement learning technology.
\item We design and implement a DDPG-based cognitive routing algorithm under the routing-oriented deep reinforcement learning theory framework.
\item We prove the preliminary feasibility and potential advantage of cognitive routing by experiments on a self-develop simulator, which is also a powerful open source tool for cognitive routing research.
\end{itemize}

The rest of our paper is organized as follows: In Section II, we introduce the definition of cognitive routing and related works. Then, we propose a routing-oriented deep reinforcement learning theory framework in Section III. Based on this framework, the design of a DDPG-based routing algorithm is descried in Section IV. In Section V, we illustrate the design of RL4Net and the implementation of the DDPG-based routing algorithm on RL4Net.  Section V is the description of experiment design and evaluation result. At last ,we conclude our work and future work in Section VI.


\section{Cognitive Routing and Related Work}

\begin{figure}[!t]
  \centering
    \includegraphics[width=3.5in]{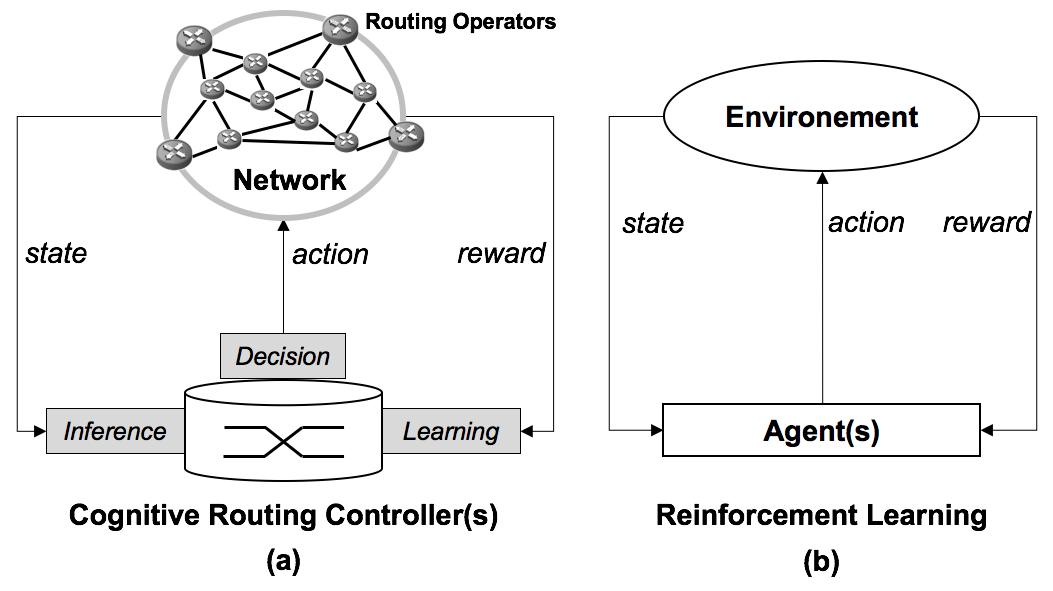}
  \caption{Cognitive Routing Framework and Reinforcement Learning\label{cognitive-RL}}
  \vspace {-10pt}
\end{figure}

Basically, the software of a router is composed by three functional components that connected by interfaces, data plane, control plane and management plane. Control plane component is responsible for exchanging routing protocols and managing routing tables. Data plane forwards data packets following routing tables produced by control plane component. For simplicity, we call the routing table managing components of control plane as \textbf{routing controller} and data plane as \textbf{routing operator.} In this paper, we focus on the core of the routing controller, routing algorithm. As we mentioned before, we can classify routing algorithms into two categories, non-quality-aware and quality-aware. Most widely used routing algorithm like RIP\cite{RIP}, IGRP\cite{IGRP}  and OSPF\cite{OSPF} are non-quality-ware. For example, the link state routing (LSR) algorithm used by OSPF choses the shortest path considering only link costs that usually related to bandwidth. This mechanism may cause congestions in heavy-load network. Although there are a number of variants like ECMP (Equal Cost Multiple Path) \cite{ECMP} attempts to decrease congestion possibilities by randomly choose a path from multiple paths with same distance. However, the defect of absence of network state information limits their improvement. To break this limitation, a number of researchers proposed to introduce quality metrics like delay, jitter and loss into parameters of routing algorithm \cite{Pragyansmita2002, Hanzo2007, Chen2007}. To deal with the complex optimization problem of increased state space, machine learning methods like Q-learning \cite{Santhi2011, Wu2013} and neural network were used to calculate candidate path for packet transmission \cite{Azzouni2017, Zhuang2019}. In this process, a rough concept of cognitive routing (not the routing algorithms only for cognitive network \cite{Cesana2011, Qadir2016}) is proposed in \cite{Jian2015} and \cite{Francois2016}. However, they did not give a clear definition of cognitive routing. Inspired by these works, we define the \textbf{cognitive routing} as: \textit{a mechanism learned from historical data for optimal routing decision by considering the inference of network quality state}. From this definition, we can see that a cognitive routing controller must have three capabilities: (1) inference network state from monitored data, (2) routing decision by considering network quality state, and (3) learning optimal routing decision policy from historical data. The architecture of cognitive routing enabled network is shown in Figure \ref{cognitive-RL}(a).

In Figure \ref{cognitive-RL}(a), if we regard the network as an environment and cognitive routing controller(s) as intelligent agent(s), the architecture of cognitive routing enabled network is similar with the reinforcement learning (RL) framework in Figure \ref{cognitive-RL}(b). Therefore, the reinforcement learning methodology is a good potential underlying methodology to implement cognitive routing controller. Actually, we are alone of thinking like this way. 
Applying RL to solve routing problem started in 1994 \cite{Boyan1994}. After this, a number of RL-based routing algorithms are proposed \cite{Mammeri2019}. However, these RL-based routing algorithms failed because of tabular-based RL method cannot handle explosive space of the combination of network state and action. In recent years, deep reinforcement learning (DRL) has been proved to be a good methodology for solving complex optimal control problem. Authors in \cite{Stampa2017} firstly used DRL in routing algorithm. After this, a small number of DRL-based routing algorithms are proposed in \cite{Mao2017, Quang2018, Xu2018, Ding2019}. Although these initial works have proved the potential of DRL for routing optimization, there are still a number of problems to be solved to achieve cognitive routing for future network.

\section{DRL Problem Definition of Cognitive Routing}

\begin{figure}[!t]
  \centering
    \includegraphics[width=3.5in]{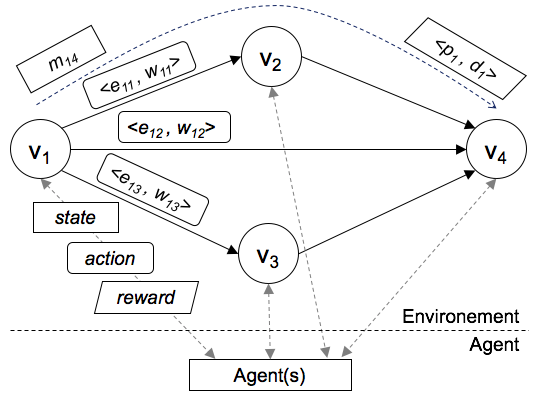}
  \caption{Sample Network Topology\label{network-topology}}
  \vspace {-10pt}
\end{figure}

We take a simple network topology shown in Figure \ref{network-topology} as an example to formulate the DRL problem of cognitive routing. Generally, a network can be denoted as $G = (V, E)$. $V$ is a set of $N$ nodes that are routers in the physical network, such as $v_1-v_4$ in Figure \ref{network-topology}. $E=\{e_{ij}\}$ is a set of directed links between nodes, which are optical fiber or copper cable between routers. If there is directed link that can sent package from router $v_i$ to $v_j$, we have $e_{ij}=1$. Otherwise, we have $e_{ij}=0$. In a period of time $T$, there are a set of $M$ packets $P=\{p_i\}$ are transmitted between routers. Each packet has an end-to-end delivery delay $d_i$, such as the delay $d_1$ of packet $p_1$ from $v_1$ to $v_4$ in Figure \ref{network-topology}. In this condition, if we have an intelligent agent (or a set of intelligent agents) than can observe the network environment and take actions on routers, we can define the factors of deep reinforcement as below:

\begin{itemize}
\item \textbf{state}: Each packet $p_i$, comes into the network via source router and departs from the network via destination router. For example, the packet $p_i$ is sent from source node $v_1$ to destination node $v_4$. For all packets $P$, we have a Traffic Matrix  $TM=\{d^t_{ij}\}$, where $d^t_{ij}$ is the sum of size of packets transmitted from $v_i$ to $v_i$ in time slot $t$. We define the state of network environment ($s^t\in S$, $S$ is the state space) as: 

\begin{equation}\label{state}
s^t=\left[\begin{array}{ccc} d^t_{11} & \hdots & d^t_{1N} \\ \vdots & \ddots & \vdots \\ d^t_{N1} & \hdots & d^t_{NN}\end{array}\right]
\end{equation}

\item \textbf{action}: Action represents how the agent change the environment. In routing context, the action of a intelligent controller is setting the routing tables of routers. Therefore, we define the action at time $t$ as the set of link weights of all nodes. Each node $v_i$ has a weight vector $W_i=<w_{11},...,w_{ij},...,w_{iN}>$, where $w_{ij}$ is the weight of link from $v_i$ to $v_j$. Then, we define the action at time $t$ ($a^t\in A$, $A$ is the action space) as: $a^t=\{W^t_1,...,W^t_i,...W^t_N\}$.

\item \textbf{reward}: Reward is the feedback information from environment to agent after agent takes an action. With different network optimization purpose, we can define different rewards. In this paper, we consider to optimize the end-to-end delay of packets delivery. Therefore, we define the reward $r^t$ as the average delay of packets in time slot $t$: $r^t=-\sum d_i / M$.

\item \textbf{policy}: Policy of agent $\pi:S\rightarrow A$ is represented by a distribution of conditional probability: $\pi(a^t|s^t)=P(A=a^t|S=s^t)$.

\end{itemize}

\begin{algorithm}[!b]
\label{algorithm-DDPG}
\caption{Cognitive Routing Algorithm based on DDPG}
\begin{algorithmic}[1]
\State Initialize online actor network $\mu$ and online critic network $Q$ with random parameters $\theta^\mu$ and $\theta^Q$, respectively. 
\State Initialize target actor network $\mu_t$ and target critic network $Q_t$ with parameters $\theta^{\mu_t}=\theta^\mu$ and $\theta^{Q_t}=\theta^Q$, respectively. 
\State Initialize a replay buffer $R$ with a Capacity $C$ and a sample threshold $T$.
\For {episode=1,...,M}
\State $s^0$ = reset($env$) 
\For {t=1,...,T}
\State $a^t = \mu(s^t)+\eta$
\State $<r^t, s^{t+1}>$ = $env$.execute($a^t$)
\State $R$.push($<s^t,a^t,r^t,s^{t+1}>$)
\If{$R$.size $>T$}
\State $B$ = $R$.sample($<s^i, a^i, r^i, s^{i+1}>$)
\State $y_i = r^i+\gamma Q^t(s^{i+1}, \mu^t (s^{i+1}))$
\State Update $\mu$ by minimizing loss function: 
\State ~~~~$L=\frac{1}{N} \sum_i (Q(s^i,a^i)-y_i)^2$
\State Update $Q$ by applying policy gradient: $\nabla_\theta^\mu J\approx$ 
\State ~~~~\small{$\frac{1}{N} \sum_i \nabla_a Q(s,a \left |\theta^Q)\right | _{s=s^i, a=\mu(s^i)} \nabla_{\theta^\mu} \mu(s \left | \theta^\mu) \right |_{s^i}$}
\State Update target networks:
\State ~~~~$\theta_{Q^t} = \tau \theta_Q + (1-\tau)\theta_{Q^t} $
\State ~~~~$\theta_{\mu^t} = \tau \theta_{\mu} + (1-\tau)\theta_{\mu^t} $
\EndIf
\EndFor
\EndFor
\end{algorithmic}
\end{algorithm}

With above definitions, we can formulate the DRL problem for cognitive routing as an \textbf{optimization} problem: how to find an optimized policy $\pi$ to maximize the reward.

\section{Design of DDPG-based Routing Algorithm}

The task of DRL agent is to optimize its policy $\pi:S \to A$ to maximize the reward. For a state $s^t \in S$ at time slot $t$, we define a value function $v^\pi(s^t)$ to evaluate the value obtained following policy $\pi$. We use a discount rate $\gamma \in [0,1)$ to decay the future rewards. $v^\pi(s^t)$ is evaluated by accumulating discounted reward as follows:

\begin{equation}
v^\pi(s^t) = E_\pi[\sum_{k=t}^\infty \gamma^{k-t} r^k(s^k, a^k)] = E_\pi[r^t(s^t,a^t)+\gamma v^\pi(s^{t+1})]
\end{equation}

We define a Q-function as: 

\begin{equation}
Q(s^t,a^t)=r^t(s^t,a^t)+\gamma E_\pi[v^\pi(s^{t+1})]
\end{equation}

An optimal policy $\pi^*$ can maximize $v^\pi(s^t)$: $v^*(s^t)=max_{a}{Q(s^t,a)}$. Therefore, the optimization problem can be solved by updating the Q-function by Temporal Difference (TD) between the target Q-value $ r^t(s^t,a^t)+\gamma max_a Q(s^{t+1},a)$ and current Q-value $Q(s^t,a^t)$ through iterative processes for all state-action pairs: $Q(s^t,a^t)=Q(s^t,a^t)+\alpha[r^t(s^t,a^t)+\gamma max_a Q(s^{t+1},a)-Q(s^t,a^t)]$, where $\alpha$ is a hyper-parameter named learning rate in the training process.

We choose widely used Deep Deterministic Policy Gradient (DDPG) algorithm \cite{DDPG} to solve the optimization problem. The designed cognitive routing algorithm based on DDPG is shown in Algorithm \ref{algorithm-DDPG}.

In Algorithm \ref{algorithm-DDPG}, line 5 resets the environment and get the initial state $s^0$ in each episode. In line 7, $\eta$ is the exploration noise, which is generated by Ornstein-Uhlenbeck process (OUProcess) \cite{Uhlenbeck1930}. In line 11, we sample a batch of $<s^i, a^i, r^i, s^{i+1}>$ tuples from replay buffer $R$. Line 13-19 is the process of update target networks off critic and actor.

\section{RL4Net and Algorithm Implementation}

\begin{figure}[!t]\label{RL4Net}
  \centering
    \includegraphics[width=3.5in]{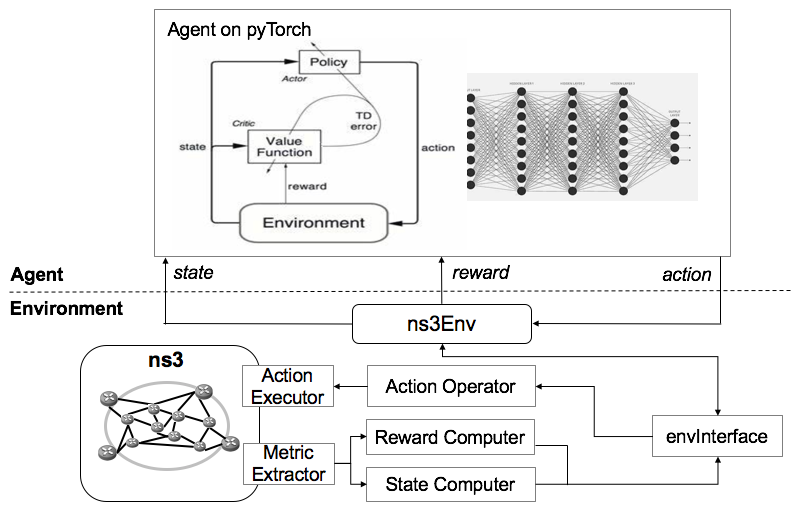}
  \caption{Architecture of RL4Net Simulator}
  \vspace {-10pt}
\end{figure}

Implementing a reinforcement learning environment and algorithms from scratch is a difficult task. Inspired by work of \cite{ns3gym}, we develop tool named RL4Net (Reinforcement Learning for Network) to facilitate the research and simulator of reinforcement learning based cognitive routing. Figure \ref{RL4Net} shows the architecture of RL4Net, which is composed by two functional blocks:

\begin{itemize}
\item \textbf{Environment}: Environment is built on widely used ns3 network simulator \cite{ns3}. We extend ns3 with six components: (1) Metric Extractor for computing quality metrics like delay and loss from ns3; (2) Computers for translating quality metrics to DRL state and reward; (3) Action Operator to get action commands from agent; (4) Action Executor for perform ns3 operations by actions; (5) ns3Env for transforming the ns3 object into DRL environment; (6) envInterface to translate between ns3 data and DRL factors. 
\item \textbf{Agent}: Agent is container of a DRL-based cognitive routing algorithm. A agent can built on various deep learning frameworks like pyTorch and Tensorflow. We implement our DDPG-based routing algorithm on pyTorch. The algorithm implementation is a python program following the logic of Algorithm \ref{algorithm-DDPG}.
\end{itemize}

Specifically, we use fully connected neural networks to implement the actor and critic of DDPG. There are four layers in actor networks, one input layer, two hidden layers and one output layer. The neuron numbers of these four layers are represented as $N_1$, $N_2$, $N_3$ and $N_4$, respectively. To scale up the action output, we multiply the output of softmax layer with a parameter $a_bound$. Network of critic is composed of three layers, one input layer, one hidden layer and one output layer. The neuron numbers of these three layers are represented as $N_5$, $N_6$ and $N_7$, respectively. Both actor and critic networks use RELU as activation function. 

\section{Experimental Evaluation}

\begin{figure}[!t]
  \centering
    \includegraphics[width=3.5in]{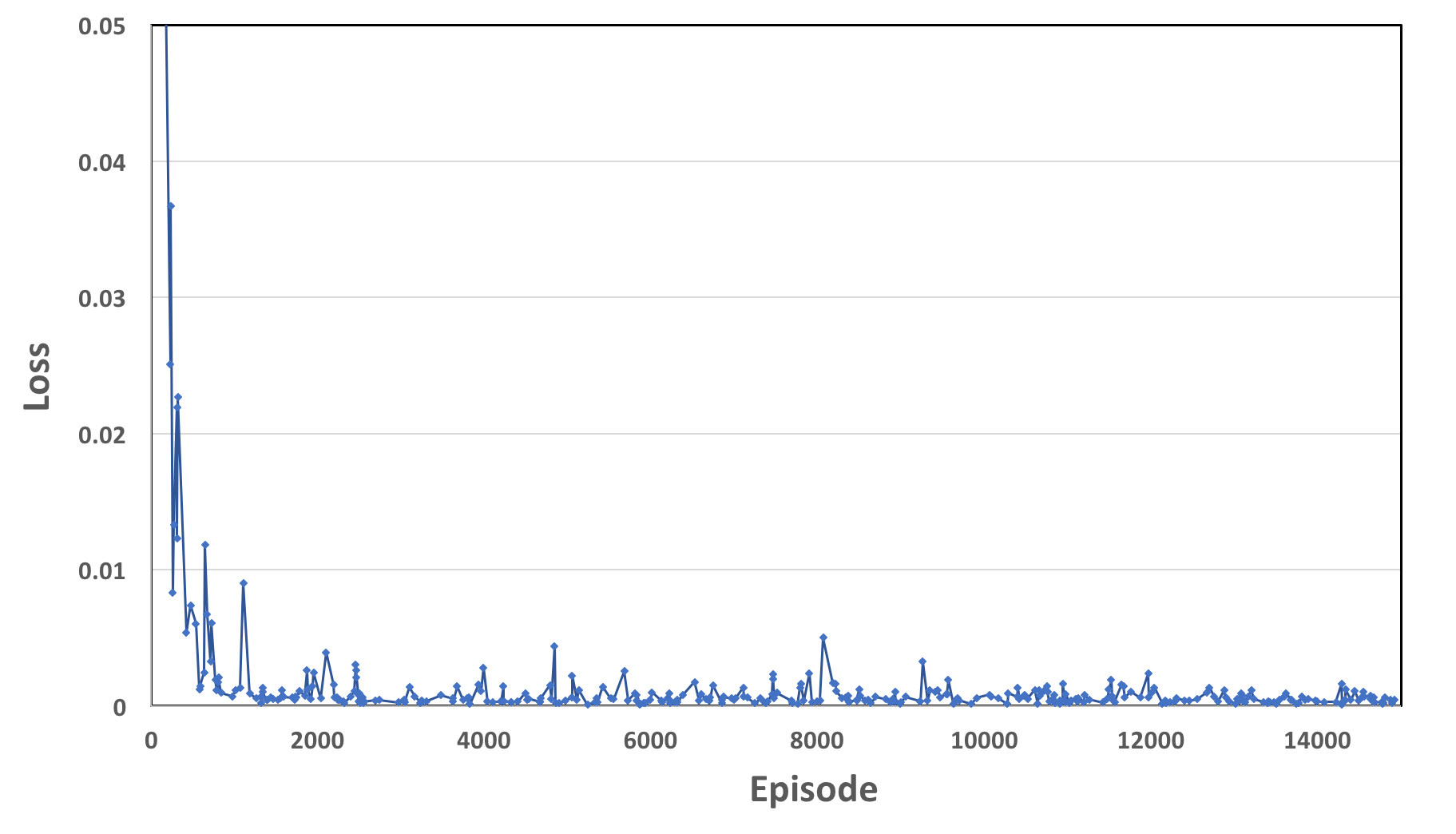}
  \caption{Loss of critic network\label{critic-loss}}
  \vspace {-10pt}
\end{figure}

\begin{figure}[!t]
  \centering
    \includegraphics[width=3.5in]{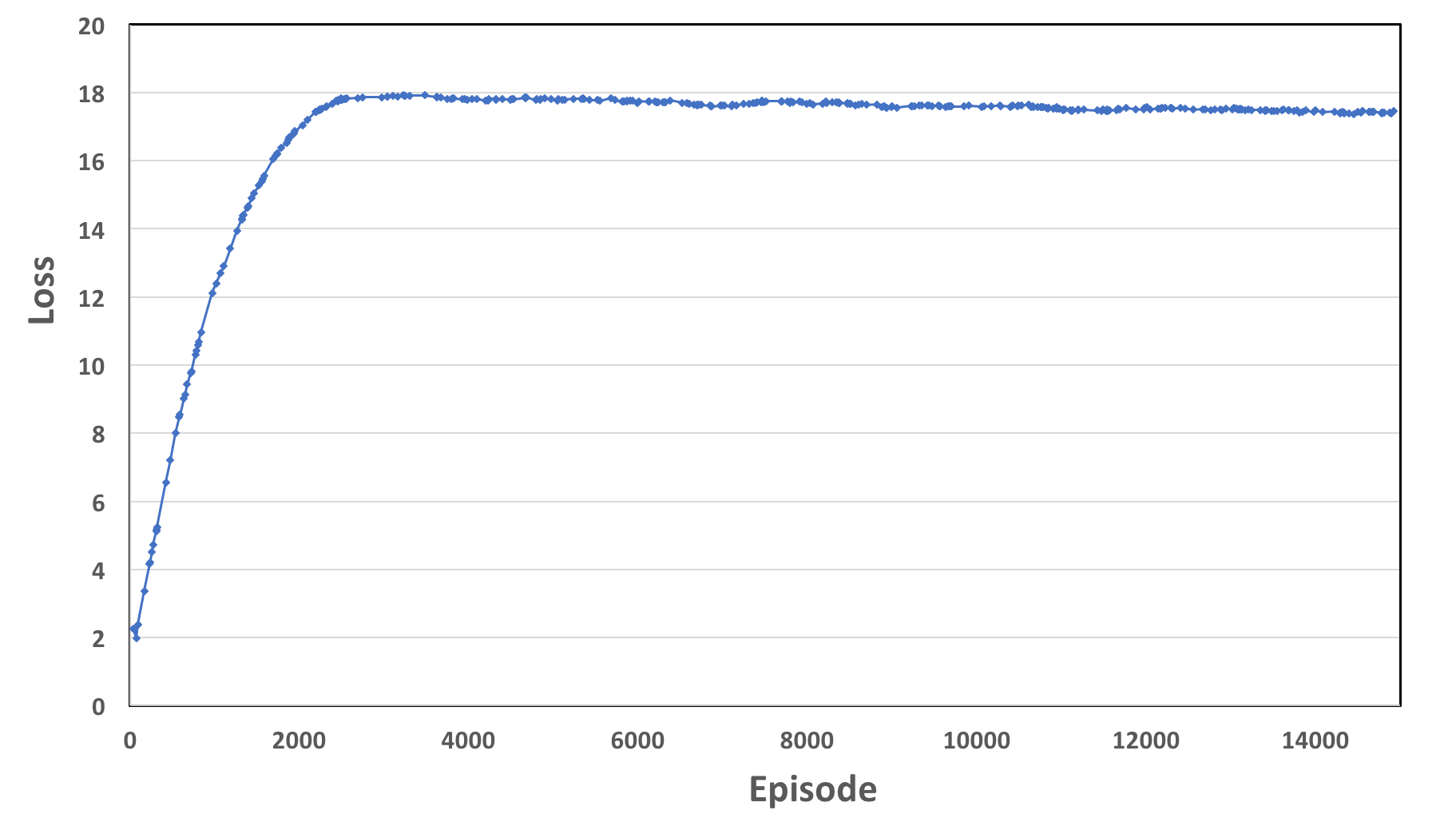}
  \caption{Loss of actor network\label{actor-reward}}
  \vspace {-10pt}
\end{figure}

\subsection{Experiment Setup}

In the experiment, we set the neuron numbers of DDPG actor and critic networks as $N_1=16$, $N_2=64$, $N_3=32$, $N_4=8$, $N_5=24$, $N_6=64$ and $N_7=1$. The scale-up parameter $a_bound$ is set to $10$. In addition, the learning rate of actor and critic, parameters $\gamma$ and $\tau$ in Algorithm \ref{algorithm-DDPG} are set to $10^{-4}$, $10^{-3}$, $0.9$ and $0.01$, respectively. The exploration noise $\eta$ is generated by parameters of $\mu=0$, $\theta=0.1$ and $\sigma=0.15$. The parameters of experience replay buffer are $C=100$, $T=64$ and $B=32$.

To evaluate our proposed DDPG-based routing algorithm, we config an experimental network topology as Figure \ref{network-topology}. Bandwidth of all links are 5Mbps. On this network, we generated a 4.636Mbps UDP flow with 1024 packet size from $v_1$ to $v_4$, which makes the link $e_{14}$ works in a heavy load condition. Under this setting, we compare the average end-to-end delivery delay of packets with other two routing algorithms, OSPF and random weight. The random weight algorithm sets the weight vector of each router randomly.

\subsection{Experiment Results}

Figure \ref{critic-loss} shows the values of loss function of critic network. As we mentioned before, the target Q-value is $y_i = r^i+\gamma Q_t(s^{i+1}, \mu_t (s^{i+1}))$. The loss function is the average square of TD-error between Q-value and its target Q-value: $L=\frac{1}{N} \sum_i(Q(s^i,a^i)-y_i)^2$. We trained the DDPG model for 43,100 steps. We can see the value of $L$ decreases with the increase of steps, which means the TD-error between Q-value and target Q-value decreases. After 15,000 step, the loss value stably remains a small value, which means the critic network is optimal enough. Therefore, we only draw values of 1-15,000 steps. 

Figure \ref{actor-reward} shows the values of loss function of actor network. We set $Q_{avg}$ as the mean of output of $Q$ network, $Q_{avg} =\frac{1}{N_7} \sum_{i=1}^{N_7} Q(i)$. The loss function of actor network is $L_a = -Q_{avg}$. We can see that the value of $L_a$ improves gradually from 1 to 16 during steps from 1-2,000. After that, $L_a$ keeps stable from 2000 step to 15,000 step, which means the actor network is successfully trained.

\begin{figure}[!t]
  \centering
    \includegraphics[width=3.5in]{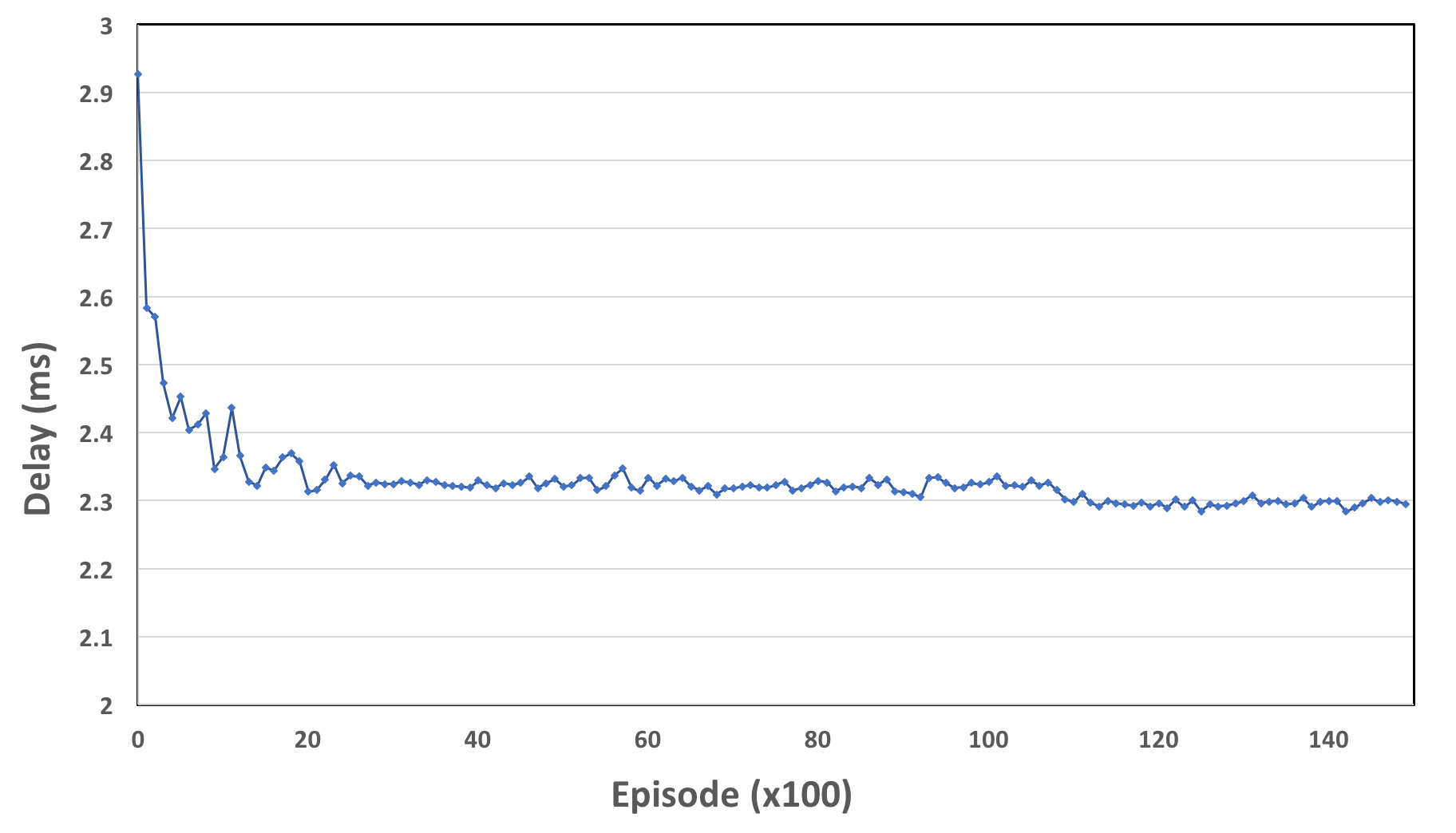}
  \caption{Average delay of training process\label{cum-delay}}
  \vspace {-10pt}
\end{figure}

Figure \ref{cum-delay} shows the average end-to-end delivery delay of packets for every 100 steps. The delay is decreased during steps 1-4,000. After that, the delay remains around 2.3ms. It shows that the DDPG algorithm has found an optimal policy.

Figure \ref{delay} shows the average delay of DDPG-based, OSPF and random weight routing algorithm. We can see that the proposed DDPG-based routing algorithm achieved the best performance with lowest end-to-end packet delivery delay after it has been trained.

\section{Conclusion and Future Work}

In this paper, we introduced a definition of cognitive routing with inference, decision and learning capabilities. Based on the definition, we proposed a deep reinforcement learning (DRL) based cognitive routing framework by defining the DRL factors in the cognitive routing environment. To facilitate the research and evaluation of DRL-based routing, we designed and developed a tool named RL4Net. A DDPG-based cognitive routing algorithm has been design and implemented on RL4Net. The experimental evaluation results showed that the proposed DDPG-based routing algorithm performs better than OSPF and random weight algorithms. Our work in this paper has proves the potential of DRL for achieving cognitive routing. In the future, we plan to extend the RL4Net to enable it configuring routers in testing network for algorithm evaluation. In addition, we will design and implement more algorithms to find effective DRL-based cognitive routing algorithm that can be used in real networks.


\end{document}